\documentstyle[fleqn,12pt]{article}
\begin{document}

\renewcommand{\theequation}{\arabic{section}.\arabic{equation}}
\renewcommand{\thesection}{\Roman{section}}
\baselineskip=28pt
\begin{titlepage}
\vspace*{2cm}

\begin{center}
{\huge   \bf          Long-Range Interaction Models\\
\vspace*{0.25cm}
                      and Yangian Symmetry}
\end{center}
\vspace*{1cm}

\begin{center}
{\bf                   Mo-Lin Ge and Yiwen Wang}\\
\vspace*{0.5cm}
{\small {\it           Theoretical Physics Division,
                       Nankai Institute of Mathematics,}}\\
{\small {\it           Tianjin 300071, People's Republic of China}}
\end{center}
\vspace*{0.3in}

\begin{abstract}
\baselineskip=28pt
The generalized Sutherland-R\"omer model and Yan models with
internal spin degree are formulated in terms of both the
Polychronakos' approach and RTT relation associated to Yang-Baxter
equation in consistent way. The {\em Yangian} symmetry is shown to
generate both the models. We finally introduce the reflection algebra
$K(u)$ to long range interaction models.\\

\end{abstract}
\end{titlepage}

\section{Introduction}
\setcounter{equation}{0}

In the last few years, a number of one-dimensional long-range interaction
models have been studied [1-10].
The typical one is Calogero-Sutherland model \cite{Cal1,Sut3},
then it is subsequently extended to the models with internal spin
degrees of freedom [5-9].
Among them an interesting approach was proposed by
Bernard-Gaudin-Haldane-Pasquier(BGHP)
who made this type of models related to the RTT relation associated with
Yang-Baxter equation(YBE) \cite{Ber1}.
The BGHP approach provides a method to deal with long-range interaction
models: for a given rational solution of YBE,
for example, $R(u)=u+P$, where $P$ is
the permutation and $u$ the spectral parameter, RTT relation gives rise
to the {\it Yangian} symmetry. With a particular realization
of the {\it Yangian}, in general, we can generate corresponding Hamiltonian
of the considered systems.

On the other hand Polychronakos had formulated the integrability in terms
of the ``coupled'' momentum operators \cite{Pol1,Min1}:
\begin{equation}
\pi_i=p_i+i\sum_{j\neq i}V_{ij}K_{ij} \label{pi}
\end{equation}
where $p_i=-i\frac{\partial}{\partial x_i}\ (\hbar=1)$,
$V_{ij}=V(x_i-x_j)$ a potential to be determined and $K_{ij}$ the
particle permutation operators. The requirements of the Hermiticity of
$\pi_i$, the absence of linear terms in $p_i$ and that only the
two-body potentials in the Hamiltonian lead to \cite{Pol1}
\begin{eqnarray}
\lefteqn{V(x) = -V(-x)\ ,} \nonumber\\
\lefteqn{H_0 \equiv \frac 1 2 \sum_i \pi^2_i=\sum_ip^2_i+\frac 1 2 \sum_{i\neq
j}
\left[ \frac{\partial}{\partial x_i} V_{ij}K_{ij}
+V^2_{ij}\right] - \frac 1 6 \sum_{i\neq j \neq k \neq i}V_{ijk}K_{ijk}}
\end{eqnarray}
where
\begin{eqnarray}
V_{ijk} &=& V_{ij}V_{jk}+V_{jk}V_{ki}+V_{ki}V_{ij} = W_{ij}+W_{jk}+W_{ki}\ ,\\
K_{ijk} &=& K_{ij}K_{jk}\nonumber
\end{eqnarray}
with $W_{ij}=W(x_i-x_j)$ being a symmetric function.
The commutation relation between $\pi_i$ and $\pi_j$ is found to be
\begin{equation}
[\pi_i,\pi_j]=\sum_{k\neq i,j}V_{ijk}(K_{ijk}-K_{jik})\ .
\end{equation}
This approach can be applied to many integrable systems, especially
to C-S model \cite{Pol1,Min1}.

Recently, Sutherland and R\"{o}mer(S-R) presented a new long-range
interaction model with the Hamiltonian \cite{Sut2}:
\begin{equation}
H_{SR}=\frac 12 \sum_i p^2_i+\sum_{i<j}l(l-1)\left[\frac{P^+_{ij}}
{sh^2x_{ij}}-\frac{P^-_{ij}}{\cosh^2x_{ij}}\right]\label{hsr}
\end{equation}
where
\begin{equation}
x_{ij}=x_i-x_j,\ P^{\pm}_{ij}=\frac{1\pm \sigma_i\sigma_j}{2}\
(\sigma^2_i=1) \label{pij}
\end{equation}
and $a,\ l$ are arbitrary parameters. Sutherland and R\"{o}mer had proved
that eq.~(\ref{hsr}) is quantum integrable. In parallel to this development
Yan proposed another model \cite{Yan1}:
\begin{equation}
H_Y=\frac 12 \sum_ip^2_i +\frac 12 \sum_{i\neq j}l
\delta (x_i-x_j)P^+_{ij} \label{hy}
\end{equation}
that was solved in terms of Bethe Ansatz. So far both the S-R model and
Yan model have not systematically been studied in terms of RTT relation.

In this paper we shall show the following points:\\
1) The models eq.~(\ref{hsr}) and eq.~(\ref{hy}) are also the
conclusion of Polychronakos'

approach.\\
2) On the basis of RTT relation the models eq.~(\ref{hsr}) and
eq.~(\ref{hy}) are related to

the realization of {\it Yangian}, namely,
they belongs to the Yang-Baxter system.

Both 1) and 2) are consistent with each other.\\
3) Further properties have been discussed that leads to other complicated con-

served quantities.

\section{Sutherland-R\"{o}mer Model and Yan Model}
\setcounter{equation}{0}

Let us first discuss the extended forms of $V_{ij}$ in eq.~(\ref{pi}) that
are different from those given by ref. \cite{Pol1,Min1}. Setting
\begin{equation}
V_{ij}=P^+_{ij}a_{ij}+P^-_{ij}b_{ij}\label{vij}
\end{equation}
where $P^{\pm}_{ij}$ are given by eq.~(\ref{pij}) and $\sigma_i$ quantum
operators obeying
\begin{eqnarray}
\sigma_iK_{ij} &=& K_{ij}\sigma_j\ ,\nonumber\\
\sigma_iK_{mn} &=& K_{mn}\sigma_i\ \ (i\neq m ,n)\ ,\nonumber
\end{eqnarray}
then by substituting eq.~(\ref{vij}) into eq.~(\ref{pi}) and doing
the parallel discussion in ref. \cite{Pol1}, we find
\begin{equation}
V_{ijk}=P^+_{ijk}A_{ijk}+P^-_{ijk}A_{ijk}+P^-_{kij}A_{kij}+P^-_{jki}A_{jki}
\end{equation}
where
\begin{eqnarray}
P^{\pm}_{ijk} &=& P^{\pm}_{ij}P^{\pm}_{ik}\ ,\nonumber\\
A_{ijk} &=& a_{ij}a_{jk}+a_{jk}a_{ki}+a_{ki}a_{ij}\ ,\nonumber\\
B_{ijk} &=& a_{ij}b_{jk}+b_{jk}a_{ki}+b_{ki}a_{ij}\ .\nonumber
\end{eqnarray}
Noting that
$P^+_{ijk}=P^+_{ikj}=\cdots=P^+_{kji}$, but $P^-_{ijk}=P^-_{jik}$ only.

The sufficient condition of the quantum integrability of
eq.~(\ref{pi}) is \cite{Pol1,Min1}
\begin{equation}
V_{ijk}=\mbox{constant} \ \ (\mbox{or \ zero}) \label{suf} \ .
\end{equation}

Now let us look for new solution of eq.~(\ref{suf})

\noindent{\bf(1)} When $A_{ijk}\neq 0$, $B_{ijk}\neq 0$,
a sufficient solution can be checked:
\begin{eqnarray}
a(x)=l\coth(ax) \ \ &&(\mbox{or} \ a(x)=l\cot(ax))\ ,\nonumber\\
b(x)=l\tanh (ax) \ \ &&(\mbox{or} \ b(x)=l\tan(ax))
\end{eqnarray}
where $x \equiv x_{ij}=x_i-x_j$, $a,\ l$ constants and
\begin{equation}
V_{ijk}=-l^2(P^+_{ijk}+P^-_{ijk}+P^-_{kij}+P^-_{jki})=-l^2\ .\label{vv}
\end{equation}
Define \cite{Min1}
\begin{equation}
H=\frac 12 \sum_i \pi^2_i-\frac {l^2}{6} \sum_{i\neq j \neq k \neq i}K_{ijk}\ ,
\end{equation}
then eq.~(\ref{vv}) leads to
\begin{equation}
H=\frac 12 \sum_i p^2_i+\sum_{i<j}l(l-aK_{ij})\left[\frac{P^+_{ij}}
{\sinh^2(ax_{ij})}-\frac{P^-_{ij}}{\cosh^2(ax_{ij})}\right]\ . \label{hrs}
\end{equation}
Eq.~(\ref{hrs}) is exactly $H_{SR}$ given by S-R \cite{Sut2} when $K_{ij}=\pm
1$.

Define
\begin{equation}
\bar {\pi}_i=\pi_i+il\sum_{i\neq j}K_{ij}\ ,
\end{equation}
then
\begin{eqnarray}
[\bar {\pi}_i, \bar {\pi}_j] & = & 2il(\bar {\pi}_i- \bar {\pi}_j)K_{ij}\\
{[ H, \pi_i]} &=& [H, \bar {\pi}_i]  =  0\ .
\end{eqnarray}
The conserved quantities are given by
\begin{equation}
I_n=\sum_i\bar{\pi}^n_i
\end{equation}
which leads to
\begin{eqnarray}
[I_n, I_m] &=& 0\ ,\label{ii}\\
{[H, I_n]} &=& 0\ ,
\end{eqnarray}
i.e. the model is quantum integrable in the sense of
Polychronakos \cite{Pol1,Min1}.

\noindent {\bf (2)} When $B_{ijk}=0$, we consider two cases\\
{\bf (a)} $A_{ijk}=0$
\begin{eqnarray}
a(x) &=& \frac l x\ ,\ \  V_{ijk}=0\ ,\nonumber\\
H &=& \frac 12 \sum_ip^2_i + \frac 12 \sum_{i\neq j}
       \frac{l(l-K_{ij})}{(x_i-x_j)^2}P^+_{ij}
\end{eqnarray}
that is well known as Calogero model
when $P^+_{ij}$ takes the value $1$.\\
{\bf (b)} $A_{ijk}=\beta^2 \neq 0$
\begin{equation}
[\pi_i, \pi_j]=\beta \sum_{k\neq i,i}P^+_{ijk}(K_{ijk}-K_{jik}) \label{pic}\ .
\end{equation}
Define
\begin{equation}
\bar {\pi}_i=\pi_i +\beta \sum_{i\neq j}P^+_{ij}K_{ij}\label{bpi}\ ,
\end{equation}
it is easy to prove that
\begin{equation}
[\bar {\pi}_i, P^+_{jk}]=0, \ \ \forall i\  \mbox{and}\ j\neq k
\end{equation}
and
\begin{eqnarray}
[\bar {\pi}_i, \bar {\pi}_j] &=& 2\beta P^+_{ij}
(\bar {\pi}_i-\bar {\pi}_j)K_{ij}\label{pi2}\ ,\\
{[ \bar {\pi}^n_i, \bar {\pi}_j]} &=& 2\beta P^+_{ij}
(\bar {\pi}^n_i-\bar {\pi}^n_j)K_{ij}\ ,
\end{eqnarray}
so that eq.~(\ref{ii}) is also satisfied. Define
\begin{equation}
H=\frac 12 \sum_i \pi^2_i+\frac {\beta^2}6 \sum_{i\neq j\neq k\neq i}
P^+_{ijk}K_{ijk}\ .
\end{equation}
With the help of eq.~(\ref{pic}), one can prove
\begin{equation}
[H, \pi_i]=[H, \bar {\pi}_i]=[H, I_n]=0\ .
\end{equation}

For the case (b) we have two sufficient solutions of $V_{ijk}$:\\
$(\mbox{b}_1)$ \hspace*{0.95cm} $a(x)=il\cot(ax)$ \  (or $a(x)=l\coth(ax)$)
\begin{eqnarray}
& & V_{ijk}=-l^2P^+_{ijk}\ ,\nonumber\\
& & H=\frac 12 \sum_i  p^2_i +\frac 1 2 \sum_{i\neq j}
\frac{l(l-aK_{ij})}{\sin^2a(x_i-x_j)}P^+_{ij}\ . \label{sc}
\end{eqnarray}
Eq.~(\ref{sc}) is the generalization of the spin chain model
considered by BGHP \cite{Ber1}.\\
$(\mbox{b}_2)$ \hspace*{0.95cm} $a(x)=l\mbox{sgn}(x)$,
\begin{equation}
\hspace*{0.8cm}H=\frac 12\sum_i p^2_i+\frac 12 \sum_{i\neq j}l(l-K_{ij})
\delta(x_i-x_j)P^+_{ij}\ .\label{hd}
\end{equation}
On condition that $K_{ij}=\pm 1$, eq.~(\ref{hd}) was first pointed out
by Yan \cite{Yan1} through Bethe Ansatz, he also found the Y-operator
defined by Yang \cite{CNY1,Bax1} for eq.~(\ref{hd})
\begin{equation}
Y^{\alpha \beta}_{ij}=\frac{1}{ik_{ij}(ik_{ij}-2c)}[ik_{ij}-c(1-
\sigma_i\sigma_j)][-ik_{ij}P^{\alpha \beta}+c(1+\sigma_i \sigma_j)]
\end{equation}
where $P$ is the permutation, $\sigma^2_i=1$ and $Y$
satisfies \cite{CNY1}
\begin{equation}
Y^{\alpha \beta}_{jk}Y^{\beta \gamma}_{ik}Y^{\alpha \beta}_{ij}=
Y^{\beta \gamma}_{ij}Y^{\alpha \beta}_{ik}Y^{\beta \gamma}_{jk}
\end{equation}
and $c=l(l\pm 1)/2$ for $K_{ij}=\pm 1$. Noting that there is only
$P^+_{ij}$ in the Hamiltonian eq.~(\ref{hd}) for the quantum
integrability.

In this section we have re-interpreted the models eq.~(\ref{hsr})
and eq.~(\ref{hy}) from the point of view of the
formulation eq.~(\ref{pi}). Next we shall set up the
{\em Yangian} \cite{Dri1} description of models eq.~(\ref{hsr})
and eq.~(\ref{hy}) through RTT relation.

\section{RTT Relation and Long-Range Interaction Models}
\setcounter{equation}{0}

Let us apply the BGHP approach \cite{Ber1} to the S-R model and Yan model.\\
The solution of Yang-Baxter equation, $R$-matrix, takes the
simplest form as
\begin{equation}
R(u)=u+\lambda P_{00^{\prime}}\label{r}
\end{equation}
and the RTT relation reads
\begin{equation}
R_{00^{\prime}}(u-v)T^0(u)T^{0^{\prime}}(v)=
T^{0^{\prime}}(v)T^0(u)R_{00^{\prime}}(u-v)\label{rtt}
\end{equation}
where $T^0(u)=T(u)\otimes 1$, $T^{0^{\prime}}=1\otimes T(u)$ and
$P_{00^{\prime}}$ is the permutation operator exchanging the two auxiliary
spaces $0$ and $0^{\prime}$. Make the expansion \cite{Ber1}
\begin{eqnarray}
T^0(u) &=& I+\sum^p_{a,b=1}X^0_{ba}\sum^{\infty}_{n=0}
           \frac{\lambda T^{ab}_n}{u^{n+1}}\ ,\label{t}\\
P_{00^{\prime}} &=& \sum^p_{a,b=1}X^0_{ba}X^{0^{\prime}}_{ab}\ .\label{p}
\end{eqnarray}
It is well known that $\{T^{ab}_n\}$ generate the {\em Yangian} \cite{Dri1}.
Substituting eqs.~(\ref{r}),~(\ref{t}) and (\ref{p}) into eq.~(\ref{rtt})
one finds
\begin{equation}
\hspace*{-0.5cm}\sum_{a,b}\sum_{cd}X^0_{ba}X^{0^{\prime}}_{dc}\sum^{\infty}
 _{n=0} \left\{ u^{-n-1}f^n_1-v^{-n-1}f^n_2 + \sum^{\infty}_{m=0}u^{-n-1}
 v^{-m-1}f^{n,m}_3 \right\} =0
\end{equation}
where
\begin{eqnarray}
f^n_1 &=& \delta_{bc}T^{ad}_n-\delta_{ad}T^{cb}_n-[T^{ab}_n,T^{cd}_0]\ ,
    \nonumber\\
f^n_2 &=& \delta_{bc}T^{ad}_n-\delta_{ad}T^{cb}_n-[T^{ab}_0,T^{cd}_n]\ ,
     \nonumber\\
f^{n,m}_3 &=& \lambda(T^{ad}_nT^{cb}_m-T^{ad}_mT^{cb}_n)+
[T^{ab}_{n+1},T^{cd}_m]-[T^{ab}_{n},T^{cd}_{m+1}]\ .\nonumber
\end{eqnarray}
For any auxiliary space $\{X_{ab}\}$ we require $f^n_1=f^n_2=f^{n,m}_3=0$.
Obviously, $f^n_1=0$ is equivalent to $f^n_2=0$. So we need only to take
\begin{equation}
f^n_1=f^{n,m}_3=0\label{f}
\end{equation}
into account.

First from $f^{n,0}_3=0$ it follows
\begin{equation}
\delta_{bc}T^{ad}_{n+1}-\delta_{ad}T^{cb}_{n+1}=\lambda (T^{ad}_0T^{cb}_n
-T^{ad}_nT^{cb}_0)+[T^{ab}_n, T^{cd}_1]
\end{equation}
which can be recast to
\begin{eqnarray}
&&T^{ad}_{n+1} = \lambda(T^{ad}_0T^{cc}_n-T^{ad}_nT^{cc}_0)+[T^{ac}_n,
   T^{cd}_1]\ (a\neq d)\ , \label{t4}\\
&&T^{aa}_{n+1}-T^{cc}_{n+1} = \lambda(T^{aa}_0T^{cc}_n-T^{aa}_nT^{cc}_0)
   +[T^{ac}_n, T^{ca}_1]\ , \label{t2}
\end{eqnarray}
where no summation for the repeating indices is taken. Eqs.~(\ref{t4}) and
(\ref{t2}) imply that $T^{ab}_n$ can be determined by iteration for
given $T^{ab}_0$ and $T^{ab}_1$.

Now let us set
\begin{eqnarray}
T^{ab}_0 &=& \sum^N_{i=1}I^{ab}_i\ , \label{ti} \\
T^{ab}_1 &=& \sum^N_{i=1}I^{ab}_iD_i \label{d}
\end{eqnarray}
and
\begin{equation}
[I^{ab}_i, I^{cd}_j]= \delta_{ij}(\delta_{bc}I^{ad}_i-\delta_{ad}I^{cb}_i)
\label{ij}
\end{equation}
where $D_i$ are operators to be determined.  Substituting
eqs.~(\ref{ti})--(\ref{ij}) into $f^1_1$ we obtain
\begin{equation}
\sum_i\sum_j I^{ab}_i[D^i, I^{cd}_j]=0\ .
\end{equation}
Further we assume
\begin{equation}
\sum_i I^{ab}_i[D_i, I^{cd}_j]=0, \mbox{for\ any}\ j \label{di}
\end{equation}
with which the $T^{ab}_2$ should satisfy
\begin{eqnarray}
\delta_{bc}T^{ad}_2-\delta_{ad}T^{cb}_2 &=& \sum_{i\neq j}I^{ab}_iI^{cd}_j
    \left\{ \lambda \sum_{k,l}I^{kl}_iI^{lk}_j (D_j-D_i)+[D_i,D_j] \right\}
    \nonumber\\
&&+\sum_i(\delta_{bc}I^{ad}_iD^2_i-\delta_{ad}I^{cb}_iD^2_i)\label{t3}\ .
\end{eqnarray}
A sufficient solution of eq.~(\ref{t3}) is
\begin{equation}
T^{ab}_2=\sum_i I^{ab}_iD^2_i
\end{equation}
with
\begin{equation}
[D_i, D_j]=\lambda \sum_{a,b}I^{ab}_j I^{ba}_i(D_i-D_j)\label{did}\ .
\end{equation}
Thus eq.~(\ref{d}) generates long-range interaction through the
eq.~(\ref{di}) and (\ref{did}).
However so far there is not simple relationship between $D_i$ and $I^{ab}_j$
which should satisfy eq.~(\ref{di}). It is very difficult to determine
the general relationship. Fortunately, BGHP \cite{Ber1} have set up the link
with the help of projection. Let the permutation groups $\Sigma_1$,
$\Sigma_2$ and $\Sigma_3$ be generated by $K_{ij}$, $P_{ij}$ and
the product $P_{ij}K_{ij}$ respectively, where $K_{ij}$ exchange
the positions of particles and $P_{ij}$ exchange the spins at position
$i$ and $j$. The projection $\rho$ was defined as
\begin{equation}
\rho (ab)=a \ \ \mbox{for}\ \forall a\in \Sigma_2, b\in \Sigma_1\ ,
\end{equation}
i.e. the wave function considered is symmetric. Let $I^{ab} _i$ be the
fundamental representations, then
\begin{equation}
P_{ij}=\sum_{a,b}I^{ab}_i I^{ba}_j\ .
\end{equation}
Suppose that there exists \cite{Ber1}
\begin{equation}
D_i=\rho ({\hat D}_i), \ \ D_i \in \Sigma_2,\ {\hat D}_i\in \Sigma_1
\label{e21}
\end{equation}
and the ${\hat D}_i$ is particle-like operators, i.e.
\begin{equation}
K_{ij}{\hat D}_i={\hat D}_jK_{ij}, \ \ K_{ij}{\hat D}_l={\hat D}_lK_{ij}
\ \  (l\neq i,j)\ .
\end{equation}
Define
\begin{equation}
T^{ab}_m=\sum_i I^{ab}_i \rho ({\hat D}^m_i) \ \ (m\geq 0)\ ,\label{tt}
\end{equation}
then\\
(a) \vspace*{-0.8cm}
\begin{equation}
[{\hat D}_j, {\hat D}_i] = \lambda \rho^{-1}(P_{ij}(D_j-D_i))
        = \lambda ({\hat D}_j-{\hat D}_i)K_{ij}\ . \label{dd}
\end{equation}
(b) $T^{ab}_m$ satisfy eq.~(\ref{f}), i.e., RTT relation eq.~(\ref{rtt}).

Actually $f^n_1=0$ is easy to be checked. By using
$$
[{\hat D}^n_i, {\hat D}^m_j] = \sum^{n-1}_{k=0}{\hat D}^k_i[{\hat D}_i,
       {\hat D}^m_j]{\hat D}^{n-k-1}_j = \lambda \sum^{n-1}_{k=0}
       {\hat D}^k_i({\hat D}^m_i-{\hat D}^m_j){\hat D}^{n-k-1}_jK_{ij}\ ,
$$
we have $f^{n,m}_3=0$.

The projection procedure is very important for it enables us to prove that
eq.~(\ref{f}) is satisfied by virtue of eq.~(\ref{e21}).

With the expansion eqs.~(\ref{t}) and the projected long-range
expansion eq.~(\ref{tt}), the hamiltonian associated to $T(u)$ is
obtained by the expansion of the deformed determinant \cite{Ber1}:
\begin{equation}
det_qT(u)=\sum_{\sigma} \epsilon(\sigma)T_{1\sigma_{1}}(u-(p-1)\lambda)
T_{2\sigma_{2}}(u-(p-2)\lambda)\cdots T_{p\sigma_{p}}(u)\ .
\end{equation}
A calculation gives
\begin{eqnarray}
det_qT(u) &=& 1+\frac {\lambda}uM+\frac{\lambda}{u^2}\left[
  \rho(\sum_i{\hat D}_i-\frac{\lambda}2 \sum_{j\neq i}K_{ij})+
  \frac{\lambda}2 M(M-1)\right]\nonumber\\
&& +\frac{\lambda}{u^3}\rho\left\{ (\sum_i{\hat D}_i-\frac{\lambda}2\sum_
  {j\neq i}K_{ij})^2+\frac{\lambda^2}{12}\sum_{i\neq j\neq k\neq i}K_{ij}K_{jk}
  \right. \nonumber\\
&& +\lambda (M-1)\sum_i ({\hat D}_i-\frac{\lambda}2\sum_{j\neq i}K_{ij})
  \nonumber\\
&& +\left.\frac{\lambda^2}6M(M-1)(M-2)+\frac{\lambda^2}4M(M-1)\right\}
   +\cdots\ .\label{det}
\end{eqnarray}
One takes the Hamiltonian as
\begin{equation}
H=\frac 12 \rho \left\{(\sum_i{\hat D}_i-\frac{\lambda}2 \sum_{i\neq j}K_
  {ij})^2+\frac{\lambda^2}{12}\sum_{i\neq j\neq k\neq i}K_{ij}K_{jk}\right\}\ .
\end{equation}
Therefore we define the Hamiltonian which have the {\em Yangian}
symmetry given by eqs.(\ref{tt}), (\ref{ij}) and (\ref{did}).
In comparison to the known models we list the expressions for ${\hat D}_i$
satisfying eq.~(\ref{dd})

\noindent{\bf (1)} \hspace*{0.5cm} $ {\hat D}_i = p_i+\frac{\lambda}2
 \sum_{i\neq j}[\mbox{sgn}(x_i-x_j)+1]K_{ij}, \ \ \lambda=2il$,
\begin{equation}
H= \frac 12 \sum_ip^2_i +\frac 12 \sum_{i\neq j}l(l-P_{ij})
      \delta(x_i-x_j)\ .\label{h1}
\end{equation}

\noindent{\bf (2)} \hspace*{0.5cm} $ {\hat D}_i = p_i+\sum_{i\neq j}l
 [i\cot a(x_i-x_j)+1]K_{ij},\  \lambda=2l$,
\begin{equation}
H = \frac 12 \sum_ip^2_i +\frac 12 \sum_{i\neq j}\frac{l(l-aP_{ij})}
{\sin^2a(x_i-x_j)}\ .\label{h2}
\end{equation}

\noindent{\bf (3)}\hspace*{0.5cm}${\hat D}_i = p_i+il\sum_{i\neq j}
[\coth a(x_i-x_j)P^+_{ij}+\tanh a(x_i-x_j)P^-_{ij}+1]K_{ij}, \ \lambda=2il\ , $
\begin{equation}
H = \frac 12 \sum_ip^2_i +\frac 12 \sum_{i\neq j}l(l-aP_{ij})
 \left(\frac{P^+_{ij}}{\sinh^2a(x_i-x_j)}-\frac{P^-_{ij}}{\cosh^2a(x_i-x_j)}
 \right).\label{h3}
\end{equation}
Eqs.~(\ref{h1}) and (\ref{h2}) were given in ref \cite{Pol1}, eq.~(\ref{h2})
was studied in ref \cite{Ber1}. Eq~(\ref{h3}) is the generalization of
S-R model.

An alternative description of transfer matrix was given by BGHP \cite{Ber1}.
Define
\begin{equation}
{\bar D}_i={\hat D}_i-\lambda \sum_{i<j}K_{ij}\ ,
\end{equation}
then
\begin{eqnarray}
[{\bar D}_i, {\bar D}_j] &=&0\ , \label{d1}\\
\left[ K_{ij},{\bar D}_k \right] &=&0\ \ (k\neq i,j)\ ,\label{d2}\\
K_{ij}{\bar D}_i-{\bar D}_jK_{ij} &=& \lambda\ .
\end{eqnarray}
It was proved that
\begin{equation}
{\bar T}_i(u)=1+\lambda \frac{P_{0i}}{u-{\bar D}_i},\ {\bar T}(u)
=\prod_i{\bar T}_i(u)\ \mbox{and}\ \rho({\bar T}(u))\label{t1}
\end{equation}
all satisfy the RTT relation.

The deformed determinant of ${\bar T}(u)$ was defined by
\begin{equation}
det_q{\bar T}(u)=\frac{\Delta_M(u+\lambda)}{\Delta_m(u)}\ ,\ \
\Delta_M(u)=\prod^M_{i=1}(u-{\bar D}_i)\ .
\end{equation}
It was proved that
\begin{equation}
\rho(det_q{\bar T}(u))=det_q(T(u))\ .
\end{equation}

To contain the model eq.(\ref{hd}), we define ${\bar D}_i$
related to the ${\bar \pi}_i$ given by eq.~(\ref{bpi}) as
\begin{equation}
{\bar D}_i={\bar \pi}_i-\beta\sum_{j<i}P^+_{ij}K_{ij}
\end{equation}
which satisfies eqs.(\ref{d1}), (\ref{d2}) and (\ref{t1}) etc. So
we can put the models eqs.~(\ref{hrs}) and (\ref{hd}) into
Yang-Baxter system.

In conclusion of this section we have shown the consistence between
{\em Yangian} symmetry and the integrability of Polychronakos for
long-range interaction models and given the interpretation of S-R model
and Yan model from the point of view of YB system.

\section{Reflection Algebra}
\setcounter{equation}{0}

The associativity of RTT relation eq.~(\ref{rtt}) is Yang-Baxter
equation(YBE) \cite{CNY1,Bax1} ($\check R(u)=PR(u)$):
\begin{equation}
{\check R}_{12}(u){\check R}_{23}(u+v){\check R}_{12}(v)=
{\check R}_{23}(v){\check R}_{12}(u+v){\check R}_{23}(u)\label{ybe}
\end{equation}
where the subscripts indicate the spaces, namely, $1\rightarrow 0,\
2\rightarrow 0^{\prime}, \ 3\rightarrow 0^{\prime \prime}$
in comparison to eq.~(\ref{rtt}).

It is well-known that for a given ${\check R}(u)$ satisfying eq.~(\ref{ybe})
there allows corresponding reflection operator $K(u)$ determined
by \cite{Skl1}
\begin{equation}
{\check R}(u-v)K_1(u){\check R}(u+v)K_1(v)=
K_1(v){\check R}(u+v)K_1(u){\check R}(u-v) \label{ref}
\end{equation}
where $K_1(u)=K(u)\otimes 1$. Eq.~(\ref{ref}) possesses the remarkable
properties \cite{Skl1}:\\
{\bf (1)} Suppose $K_{\pm}(u)$ are c-number solutions of eq.~(\ref{ref}),
so do ${\tilde K}_{\pm}(u)$
\begin{equation}
{\tilde K}_{\pm}(u)=T(u)K_{\pm}(u)T^{-1}(-u)\ . \label{kt}
\end{equation}
{\bf (2)} Define
\begin{equation}
t(u)=\mbox{tr}[K_+(u+\lambda)T(u)K_-(u)T^{-1}(-u)]\ ,
\end{equation}
then
\begin{equation}
[t(u), t(v)]=0 \label{tc}\ ,
\end{equation}
i.e. $t(u)$ forms a commuting family. In order to solve $K(u)$
in eq.~(\ref{ref}) we make expansion:
\begin{equation}
K_0(u)=\sum_{a,b}\sum_n X^0_{ab}K^{(n)}_{ab}u^{-n}\ .\label{k0}
\end{equation}
Substituting eq.~(\ref{k0}) into eq.~(\ref{ref}) after calculations
one obtains
\begin{eqnarray}
\hspace*{-1.6cm}F^{n,m}_{ab,cd} &=& \delta_{bc}[K^{(n)},K^{(m)}]_{ad}
    +\delta_{ac}\sum_e(K^{(n+1)}_{be}K^{(m)}_{ed}+K^{(n)}_{be}K^{(m+1)}_{ed})
    \nonumber\\
&& +\delta_{bd}\sum_e(K^{(m+1)}_{ae}K^{(n)}_{ec}-K^{(n)}_{ae}K^{(m+1)}_{ec})
    +[K^{(n+2)}_{bc}, K^{(m)}_{ad}]-[K^{(n)}_{bc}, K^{(m+2)}_{ad}]
    \nonumber\\
&& +\left[K^{(n+1)}_{ac}K^{(m)}_{bd}-K^{(m)}_{ac}K^{(n+1)}_{bd}+
    K^{(n)}_{ac}K^{(m+1)}_{bd}-K^{(m+1)}_{ac}K^{(n)}_{bd} \right] \\
&=& 0 \ .
\end{eqnarray}
It follows
\begin{equation}
[K^{(0)}_{ab}, K^{(m)}_{cd}]=0\ .
\end{equation}
Suppose $K^{(0)}_{ab}=\delta_{ab}$, the iteration relation reads
\begin{eqnarray}
\hspace*{-1.2cm}\delta_{bd}K^{(m+2)}_{ac}-\delta_{ac}K^{(m+2)}_{bd} &=&
   \frac 12 \left \{\delta_{ac}[K^{(2)}, K^{(m)}]_{bd}-\delta_{bd}[K^{(1)},
   K^{(m+1)}]_{ac} \right.\nonumber\\
&& +K^{(2)}_{ac}K^{(m)}_{bd}-K^{(m)}_{ac}K^{(2)}_{bd}
   +K^{(1)}_{ac}K^{(m+1)}_{bd}-K^{(m+1)}_{ac}K^{(1)}_{bd} \nonumber\\
&& +\left.\delta_{bc}[K^{(1)},K^{(m)}]_{bd}+[K^{(3)}_{bc},K^{(m)}_{ad}]
   \right\} \ (m>1)\ . \label{it}
\end{eqnarray}
Eq.~(\ref{it}) tells that $K^{(m)}$ can be found if $K^{(1)}$, $K^{(2)}$
and $K^{(3)}$ are given properly.

Now let us consider the simplest case where $K(u)$ is a $2\times 2$
matrix given by eq.~(\ref{ks})(see below). Denote
\begin{eqnarray}
T(u)= \left[ \begin{array}{cc}T_{11}(u) & T_{12}(u)\\
      T_{21}(u) & T_{22}(u)\end{array} \right]\ ,
\end{eqnarray}
then
\begin{eqnarray}
T^{-1}(u)=[det_qT(u)]^{-1} \left[ \begin{array}{cc}T_{22}(u-\lambda) &
-T_{12}(u-\lambda)\\-T_{21}(u-\lambda) & T_{11}(u-\lambda)\end{array}
\right]\ .\label{inv}
\end{eqnarray}
Since $det_qT(u)$ commutes with $T_{ab}(v)$ one does not care the common
factor appearing in eq.(\ref{inv}). We consider the simplest case when
$K_{\pm}=1$ and denote
\begin{equation}
K(u)=T(u)T^{-1}(-u)\ . \label{ks}
\end{equation}
Now let us see what happens for the long-range interaction model where
$T(u)$ is given by eq.(\ref{tt}).  Noting that
\begin{eqnarray}
K_{11}(u)&=&T_{11}(u)T_{22}(-u-\lambda)-T_{12}(u)T_{21}(-u-\lambda)\
,\nonumber\\
K_{12}(u)&=&T_{12}(u)T_{11}(-u-\lambda)-T_{11}(u)T_{12}(-u-\lambda)\
,\nonumber\\
K_{21}(u)&=&T_{21}(u)T_{22}(-u-\lambda)-T_{22}(u)T_{21}(-u-\lambda)\
,\nonumber\\
K_{22}(u)&=&T_{22}(u)T_{11}(-u-\lambda)-T_{21}(u)T_{12}(-u-\lambda)\
.\label{kkk}
\end{eqnarray}
The $T_{ab}(u)$ in eq.~(\ref{kkk}) can be expanded in the terms of
eqs.~(\ref{t}) and (\ref{tt}) which give the $T_{ab}(u)$:
\begin{equation}
T_{ab}(u)=\delta_{ab}+\lambda\sum_iI^{ba}_id_i(u)\label{dt}
\end{equation}
where $d_i(u)=\rho(\frac{1}{u-{\hat D}_i})$. Substituting eq.~(\ref{dt})
into eq.~(\ref{kkk}) we find
\begin{eqnarray}
\hspace*{-1.5cm}
K_{11}(u)&=&1+\lambda\sum_i[I^{11}_id_i(u)+I^{22}_id_i(-u-\lambda)-
   \lambda I^{22}_id_i(u)d_i(-u-\lambda)]\nonumber\\
&& +\lambda^2\sum_{i\neq j}(I^{11}_iI^{22}_j-I^{21}_iI^{12}_j)
   d_i(u)d_j(u-\lambda)\ ,\nonumber\\
K_{12}(u)&=&\lambda \sum_II^{21}_i[d_i(u)-d_i(-u-\lambda)+\lambda d_i(u)
   d_i(-u-\lambda)] \nonumber\\
&& +\lambda^2\sum_{i\neq j}(I^{21}_iI^{11}_j-I^{21}_jI^{11}_i)d_i(u)
   d_j(-u-\lambda)\ ,\nonumber\\
K_{21}(u)&=&\lambda \sum_II^{12}_i[d_i(u)-d_i(-u-\lambda)+\lambda d_i(u)
   d_i(-u-\lambda)] \nonumber\\
&& +\lambda^2\sum_{i\neq j}(I^{12}_iI^{22}_j-I^{12}_jI^{22}_i)d_i(u)
   d_j(-u-\lambda)\ ,\nonumber\\
K_{22}(u)&=&1+\lambda\sum_i[I^{22}_id_i(u)+I^{11}_id_i(-u-\lambda)-
   \lambda I^{11}_id_i(u)d_i(-u-\lambda)]\nonumber\\
&& +\lambda^2\sum_{i\neq j}(I^{22}_iI^{11}_j-I^{12}_iI^{21}_j)
   d_i(u)d_j(u-\lambda) \label{kn}
\end{eqnarray}
and
\begin{eqnarray}
\hspace*{-1.5cm}
t(u)&=&K_{11}(u)+K_{22}(u)\nonumber\\
&=& 2+\frac {\lambda}{u^2}[2\sum_iD_i+\lambda\sum_{i\neq j}P_{ij}+C_1]
    +\frac{\lambda^2}{u^3}[\sum_iD_i+\lambda \sum_{i\neq j}P_{ij}+C_2]
    \nonumber\\
&& +\frac{\lambda}{u^4}\sum_i\left\{ 2\rho ({\hat D}_i+\frac{\lambda}2)^3-
    2(N-1)\lambda D_i-(N-1)\lambda^2 D_i\right. \nonumber\\
&& +\lambda\sum_{j\neq i}\rho({\hat D}_i{\hat D}_j)+ 2\lambda
    \sum_{j\neq i}P_{ij}\rho({\hat D}^2_i)+\lambda^2\sum_{j\neq i}
     P_{ij}D_i\nonumber\\
&& +\left. \lambda^3\sum_{j\neq i}P_{ij}-\lambda \sum_{j\neq i}P_{ij}
   \rho({\hat D}_i{\hat D}_j)\right\}+0(u^{-4})\label{th}
\end{eqnarray}
where $C_1$ and $C_2$ are constants. Obviously the second term commutes
with the third one on the RHS of eq.~(\ref{th}).

Here we would like to emphasize that the $t(u)$ does not generate conserved
quantities.

The physical meaning of eq.~(\ref{th}) for the long-range interaction models
is not clear yet. It deserves more knowledge in this area to be explored.
What we would like to say is that the simplest form of reflection matrix
$K(u)$ for long-range interaction models can really be calculated.
Substituting variety of the forms of $\hat D_i$ given in section 3, the
reflection matrix $K(u)$ can explicitly be expressed by the interactions.

We would like to thank Profs. F.D.M. Haldane, B. Sutherland,
Y.S. Wu, M. Wadati and Yan for valuble discussions.
This work was supported in part by the National
Natural Science Foundation of China.

\end{document}